# Comprehensive evaluations of a Prototype Full Field-of-View Photon Counting CT System through Phantom Studies

Xiaohui Zhan[1,*], Ruoqiao Zhang[1], Xiaofeng Niu[1], Ilmar Hein[1], Brent Budden[1], Shuoxing Wu[1], Nicolay Markov[1], Cameron Clarke[1], Yi Qiang[1], Hiroki Taguchi[2], Keiichi Nomura[3], Yoshihisa Muramatsu[3], Zhou Yu[1], Tatsushi Kobayashi[3], Richard Thompson[1], Hiroaki Miyazaki[2], Hiroaki Nakai[2]

[1]Canon Medical Research USA, Inc., 706 Deerpath Drive, Vernon Hills, IL, USA 60061
[2]Canon Medical System Corporation, Otawara, Tochigi, Japan
[3]National Cancer Centre Hospital East, 6-5-1 Kashiwanoha, Kashiwa, Japan

## ABSTRACT

***Objective*** Photon counting CT (PCCT) has been a research focus in the last two decades. Recent studies and advancements have demonstrated that systems using semiconductor-based photon counting detectors (PCDs) have the potential to provide better contrast, noise and spatial resolution performance compared to conventional scintillator-based systems. With multi-energy threshold detection, PCD can simultaneously provide the photon energy measurement and enable material decomposition for spectral imaging. In this work, we report a performance evaluation of our first CdZnTe-based prototype full-size photon counting CT system through various phantom imaging studies. ***Approach*** This prototype system supports a 500 mm scan field-of-view (FOV) and 10 mm cone coverage at isocenter. Phantom scans were acquired using 120 kVp from 50 to 400 mAs to assess the imaging performance on: CT number accuracy, uniformity, noise, spatial resolution, material differentiation and quantification. ***Main Results*** Both qualitative and quantitative evaluations show that PCCT has superior imaging performance with lower noise and improved spatial resolution compared to conventional energy integrating detector (EID)-CT. Using projection domain material decomposition approach with multiple energy bin measurements, PCCT virtual monoenergetic images (VMIs) have lower noise, and superior performance in quantifying iodine and calcium concentrations. These improvements lead to increased contrast-to-noise ratio (CNR) for both high and low contrast study objects compared to EID-CT. PCCT can also generate super-high resolution (SHR) images using much smaller detector pixel size than EID-CT and dramatically improve image spatial resolution. ***Significance*** Improved spatial resolution and more accurate material quantification with reduced image noise on PCCT can potentially lead to better diagnosis at reduced radiation dose compared with conventional EID-CT. Increased CNR achieved by PCCT suggests potential reduction in iodine contrast media load, resulting in better patient safety and reduced cost.

---

[*] Send correspondence to Xiaohui Zhan <xzhan@mru.medical.canon>

1. **INTRODUCTION**

New CT systems using semiconductor-based photon counting detectors (PCDs) are under active research and development [1][2][3][4][5] . As an emerging technology, PCDs have been introduced for experimentation and prototyping since decades ago. Recently, the first PCCT clinical product was released and marks a new era for wider clinical adoption [6]. Many studies have demonstrated its superior performance and the associated benefits in various clinical applications to conventional energy-integrating detector (EID) CT [4][7][8][9][10][11][12]. For conventional scintillator-based EID, the measurement needs a two-step conversion process: the absorbed photon energy is first converted to optical photons, and then optical photons are converted to electrical signals through a photodiode [13]. As a result, the amplitude of the signal is proportional to the photon energy, and the lower energy photons that have more material resolving power are downweighed. Another limitation from such an energy integrating process is that the electronic noise from the front-end electronics will always be part of the measurement. When the number of photons is low, the electronic noise becomes dominant and gradually degrades the image quality. In addition, between the scintillator pixels, a reflector of finite thickness is needed to prevent optical crosstalk, which increases the dose penalty when the pixel size gets small. All these limitations in EID can be largely mitigated or resolved in photon counting detector (PCD), for which the absorbed photon energy is converted to an electric signal directly. Instead of measuring the total energy deposited in each time interval for EID, PCD measures the energy of each incident photon [5][9]. This type of measurement results in an equal weighting of the transmitted photons and effectively removes the front-end electronic noise by setting a proper triggering threshold. The whole detector is active without the need for reflective material between pixels and allows for smaller pixel size without dose penalty. With multiple energy bin measurements, spectral information can be simultaneously obtained for spectral imaging which is mainly achieved by the current dual energy (DE) EID-CT [13][14], either by using a dual-source, fast kVp switching or dual-layer EID technology. Moreover, with a flexible energy threshold setting, PCD allows measurements to target specific K-edge high Z materials and enables K-edge imaging from common contrast agents such as iodine and gadolinium to novel nanoparticles [15][16][17][18][20].

For the detector materials, CdTe and CdZnTe are the most common room temperature semiconductors for such applications, benefiting from their wide band gap, high density, and high effective atomic number [19]. CdTe has a longer history being applied on such applications but typically suffers from a higher dark current and is more prone to polarization, as compared to CdZnTe. Both materials face challenges with regards to manufacturing (brittleness) and performance (e.g., thermal stability, charge trapping). Furthermore, due to the nature of the detection physics processes in these semiconductor materials, the detector spectral response usually degrades from ideal because of effects such as x-ray florescence (k-escape), Compton scattering, charge sharing, as well as pulse pileup and other complications from the associated front-end electronics [5][21]. In order to assess the practical benefits of PCCT compared to conventional EID-CT with all these non-ideal factors and explore potential new applications, a CdZnTe-based full-size photon counting CT prototype system has been built and studied.

In this manuscript, we first introduce the key technical aspects of our engineering prototype photon counting CT system, followed by a series of phantom studies for a comprehensive assessment of its imaging performance. Example images from this prototype system associated with qualitative and quantitative analysis results are presented and discussed. In particular, some of the key studies are directly compared with a Canon conventional EID-CT to demonstrate the differences introduced by PCD. In the end, we give a brief summary and outlook to the future system development plan.

## 2. MATERIALS AND METHODS

**System Description**

The engineering prototype PCCT system is built based on a Canon Aquilion ONE ViSION CT [22] gantry. The CdZnTe-based photon counting detector array populates the full 500 mm FOV in the fan angle and covers up to 9.92 mm in the z-direction at isocenter. A 1-D anti-scatter grid (ASG) is placed on top of the detector plane to reduce scattered photons. As illustrated in Figure 1, for normal resolution (NR) mode, the readouts of a 3×3 grid of micro pixels are summed as input for image reconstruction, which produces the same in-plane detection pitch as that of Canon's conventional EID-CT. For super high resolution (SHR) mode, the readout of each micro pixel is used for processing and image reconstruction. Each micro pixel can output up to six energy bins of measurements starting from 20 keV. The counting mode generates images based on the events with photon energy greater than 30 keV, and the spectral mode generates images using 5 closed energy bins with threshold settings of 30/45/55/65/80 keV, respectively.

After the electrical charges are induced in the CdZnTe sensor from the incident photon, the signal goes through a series of processing in the photon counting application specific integrated circuit (ASIC) and generates the counting measurements. The ASIC employs a charge-sensitive amplifier (CSA) and related circuitry design to allow fast signal triggering. Such readout is capable of a uniformly-distributed maximum event rate defined by the inverse of the elapsed time (deadtime), and therefore follows a non-paralyzable detection model [21]. Optimal deadtime enhances maximum count rate while also allowing for sufficient collection of the induced charge. A typical counting curve for a pre-defined deadtime is shown in Figure 2, where the measured output count rates (OCR) are plotted against the estimated incident count rates (ICR). The integrated voltage signal for each pixel is filtered and fed to six energy discriminators where it is compared to pre-programmed reference levels and appropriate counters incremented.

The technical specifications of this engineering prototype together with the EID system used for comparison are given in Table 1. In this study, scans were all acquired in circular mode at 120 kVp with 1 second per rotation speed and tube current from 50 to 400 mA.

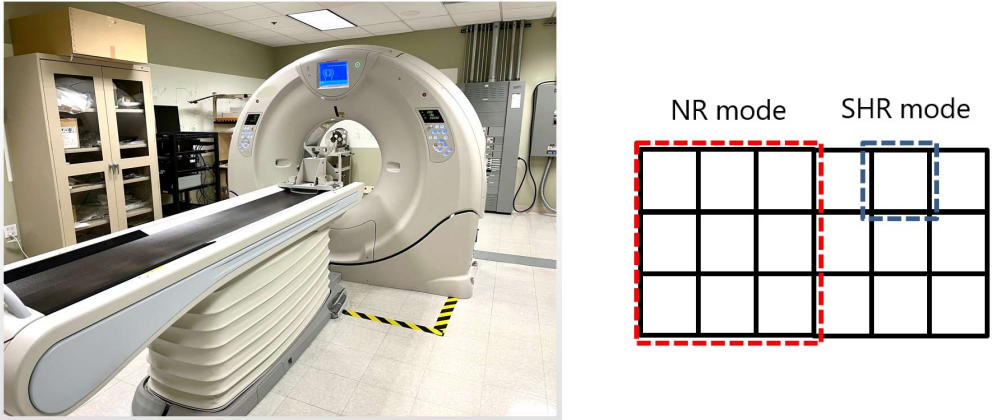

Figure 1: (Left) Canon's first engineering prototype PCCT system. (Right) An illustration of the prototype PCCT detector pixel readout schematic: the normal resolution (NR) mode combines the readout of a 3×3 micro pixels, and the SHR mode utilizes the readout of individual micro pixels for processing and image reconstruction.

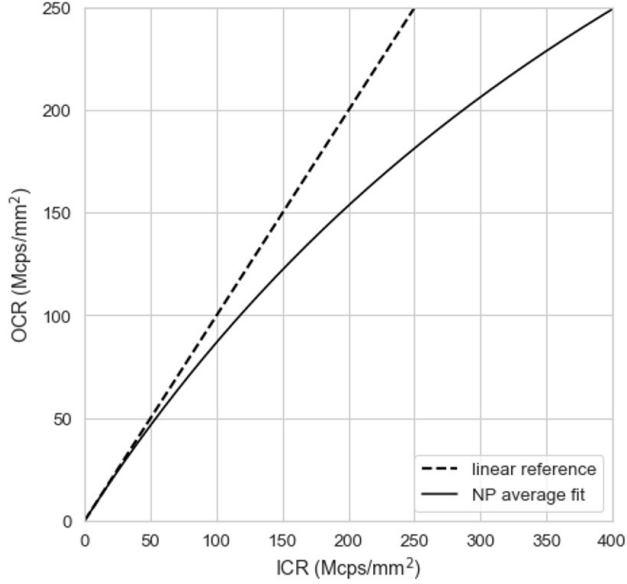

Figure 2: Typical PCD counting curve (as in mega-count-per-second) with a pre-programmed deadtime. It can be seen that the ASIC readout approximately follows a non-paralyzable detection model (solid line). When the incident count rate (ICR) increases, the output count rate (OCR) gradually deviates from the ideal linear reference (dash line) due to pulse pileup.

Table 1: Prototype PCCT system technical specification compared with a Canon EID-CT system

|  | **Technical specification overview** |  |
| --- | --- | --- |
| System | PCCT Engineering Prototype | EID-CT |
| Platform | Canon Aquilion ONE ViSION | |
| Detector material | CdZnTe | GOS + PDA |
| Collimation | 16×0.62 mm (NR), 48×0.21 mm (SHR) | 80×0.5 mm |
| Scan FOV | 500 mm | |
| Tube voltage | 120 kVp | |
| Tube current | 50/100/200/400 mA | |
| Focal Size | Large (1.6 mm×1.5 mm), Small (0.9 mm×0.8 mm) | |
| Rotation speed | 1 s | |
| Scan mode | Circular | |
| Readout mode | NR (3×3), SHR (1×1), 6 energy bins (20/30/45/55/65/80 keV) | NR |
| Recon/kernel | FBP, FC13 (Standard Body) | |

The high-level data processing flow diagram is illustrated in Figure 3. For this prototype system, we configured it in a way that it always outputs data in micro pixel mode with all six energy bins to allow in-depth studies.

After scans, the measured projection data, energy bin counts $N_i$, first went through a few preprocessing steps including tube flux variation correction and data reformation. For the counting mode, counts with photon energy above 30 keV were summed for the counting line-integral sinogram estimation. During this step, the beam hardening correction was applied to account for the effect of the polychromatic beam spectrum. The counting line-integral sinogram then went through a filtered back-projection (FBP) reconstruction and some postprocessing steps to generate the counting image. For the spectral mode, five energy bin counts $N_i$ ($i = 2,...,6$) were used for a projection-domain material decomposition to generate two basis material (e.g., water/bone) pathlength sinograms [23][24][25]. The two basis material pathlength sinograms then went through an FBP reconstruction to generate the basis material images. The basis material images were then synthesized to generate the virtual monoenergetic images (VMIs) [26]. The iodine or calcium map which contains concentrations of iodine or calcium was generated by using 60/90 keV VMI pair with an in-house software tool, which was calibrated using ground-truth iodine or calcium concentration values.

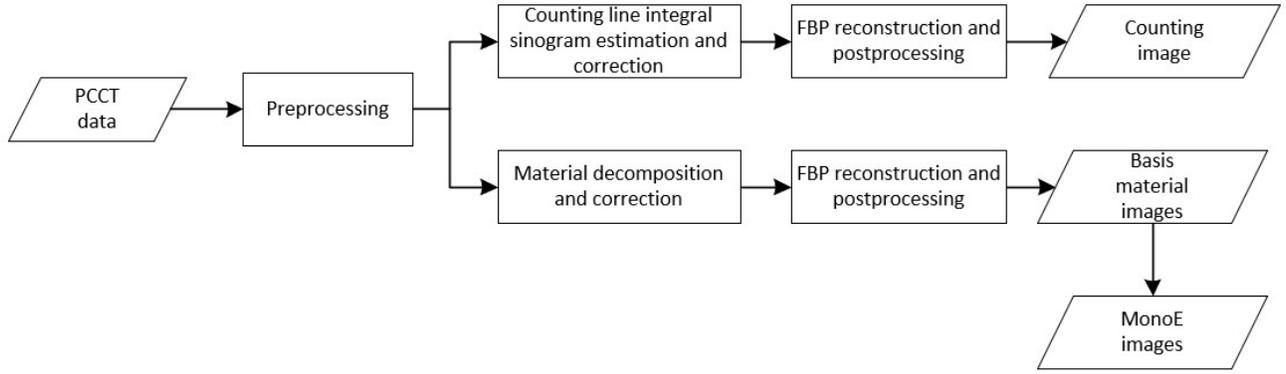

Figure 3: A high-level PCCT data processing flow diagram to generate counting and spectral images.

One of the major challenges in generating good image quality for PCCT is to accurately calibrate the detector response and establish a correct forward model [27][28][29][30]. In order to correct for the pixel-to-pixel detector response variation, each pixel needs a unique response calibration table as described in a generalized forward model below:

$$N_i(i = 1,...,n) = \int_{T_i}^{T_{i+1}} \Phi_i(E') \int_{Emin}^{Emax} N_0 S_0(E) D(E, E') e^{\int \mu L} dE dE'$$

$$\Phi_i = \begin{cases} 1, & T_i \leq E < T_{i+1} \\ 0, & others \end{cases}.$$

$\Phi_i$ stands for the PCD ideal bin response function. $E$ is the measured photon energy, and $E'$ is the incident photon energy. $S_0(E)$ and $D(E, E')$ represent the incident beam spectrum and detection response function, respectively. Their product was calibrated through a set of known attenuation samples. $Emin$ and $Emax$ define the energy range of the input spectrum. $T_i$ and $T_{i+1}$ define the energy thresholds of each energy bin. In order to achieve optimized image quality, a small number of mal-functioning pixels were excluded in the data processing.

**Phantom Studies and Image Quality Evaluations**

After a series of system and detector related calibrations, several phantoms were scanned at multiple dose levels and images were reconstructed for comprehensive evaluations on the following aspects:

1) CT number accuracy and uniformity
2) Spatial resolution and noise
3) Material quantification accuracy based on VMIs and contrast-to-noise ratio (CNR)

Through these evaluations, we explored the benefits of PCD measurements by acquiring the same phantom scans from a Canon EID-CT system (Aquilion ONE ViSION)[22]. For counting image comparison, EID-CT single-energy scans were acquired at matched radiation doses with PCCT scans. For spectral image comparison, EID-CT dual-energy scans were collected using a sequential (rotate-rotate) dual-energy acquisition with 80/135 kVp pair [14]. The mA for either kVp in the dual-energy mode was set so that the total radiation dose (CTDIvol) of the dual-energy scan matched the dose of PCCT and EID-CT single-energy scans. With the same gantry geometry, X-ray tube, beam filtration, and image reconstruction kernels (FC13) at matched radiation doses, one can perform a rigorous comparison between PCCT and EID-CT systems, and demonstrate the improvements introduced solely by the detection technology and the associated changes.

**CT number accuracy, uniformity, and noise**

CT number accuracy and uniformity were assessed using 1) a set of water cylindrical phantoms with diameters of 18/24/32/40 cm; and 2) a 32 cm diameter Canon TOS phantom with multiple material inserts (Figure 4). These phantoms were scanned with 120kVp, 50/100/200/400 mAs. PCCT images were reconstructed with 5 mm slice thickness, and 512x512 matrix size using FBP. According to IEC standards [31], the center regions of interest (ROIs) of the water phantoms, of which the diameter was 40% of that of the underlying water phantoms (Figure 15), were selected for the CT number measurements at different dose levels. For the Canon

TOS phantom, ROIs at each of the material (water, Delrin, Acryl, 66Nylon, Polypropylene (P.P.)) were measured for CT numbers and compared with EID-CT reference values. The center and 4 peripheral ROIs of the water phantoms were measured for CT number uniformity assessment.

For image noise comparison, the water phantoms were also scanned on the EID-CT system using same scanning and reconstruction parameters as used on the PCCT system. The centered ROI used for water CT number measurement was also used for image noise measurement. In addition to directly comparing PCCT and EID-CT, we also compared the image noise between PCCT and EID-CT with matched Modulation Transfer Function (MTF). This was achieved by applying a carefully tuned 2-D Gaussian filter on the PCCT sinogram data to match the MTF of EID-CT.

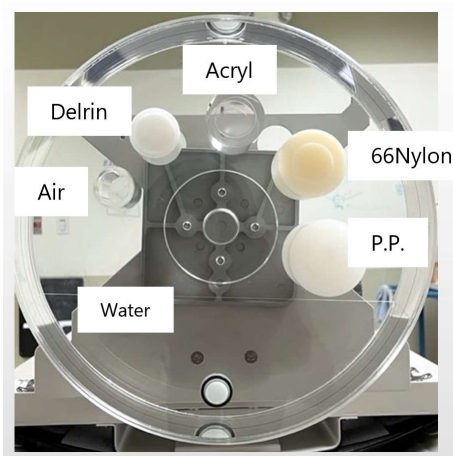

Figure 4: Canon TOS system phantom with a diameter of 32cm. It has inserts of different materials for CT number accuracy evaluation.

**In-plane spatial resolution**

To evaluate the in-plane resolution on the normal-resolution counting and spectral modes, a Catphan CTP682 module (The Phantom Laboratory, NY, USA) (Figure 5 left) was scanned on both PCCT and EID-CT systems with the same scan parameters (120kVp, 400mAs, large focal spot). Both PCCT and EID-CT images were reconstructed with 5 mm slice thickness, display FOV 60 mm, and 512x512 matrix size using FBP. The radial edge profile of the Teflon pin was used to calculate the MTF.

Three phantoms were used to compare the in-plane resolution of counting images in NR and SHR modes on the PCCT system, namely Catphan CTP682, Catphan CTP714 (Figure 5 right), and LUNGMAN phantom (Figure 6) (Kyoto Kagaku Co. Ltd., Japan). For this comparison, all three phantoms were scanned using 120kVp, 200mAs, and small focal spot. All images were reconstructed to 1024x1024 image size. For Catphan CTP628 module, the display FOV was 60 mm and the slice thickness was 5 mm. For Catphan CTP714 module, the display FOV was 120 mm and the slice thickness was 5 mm. For LUNGMAN phantom, the display FOV was 300 mm and the slice thickness was 0.62 mm. Among the three phantoms, the Teflon pin on Catphan CTP628 was used for MTF measurement, while the high contrast line pairs on Catphan CTP714 and the lung tissue structure on LUNGMAN phantom were used for a visual inspection.

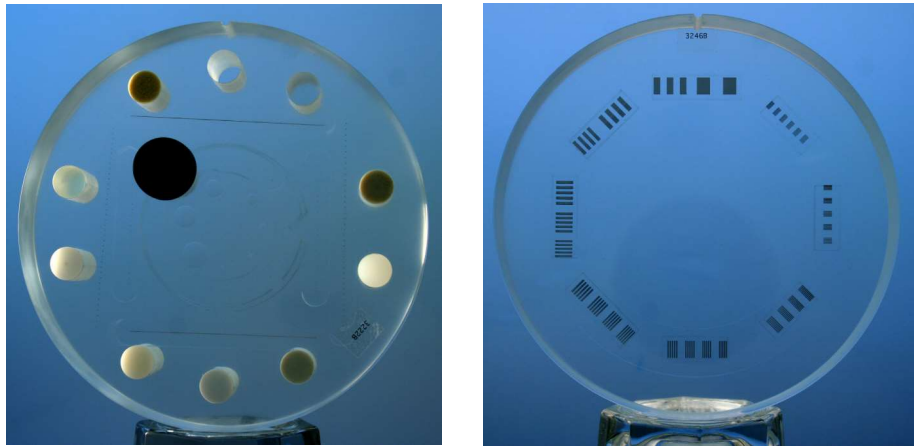

Figure 5: Catphan CTP682 (left) and CTP714 (right) (The Phantom Laboratory, NY, USA) used for image in-plane spatial resolution evaluation.

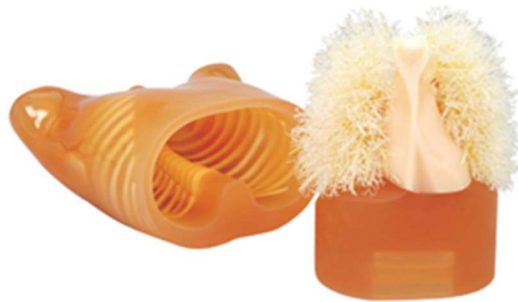

Figure 6: LUNGMAN phantom (Kyoto Kagaku Co. Ltd., Japan) was used for in-plane spatial resolution inspection.

**Material decomposition accuracy**

To examine material decomposition accuracy in spectral mode, a Multi-energy CT phantom (Sun Nuclear Corporation, WI, USA) of head size (20 cm diameter) (Figure 7 left) was used for evaluations. Iodine inserts with concentration of 2/5/10/15 mg/ml and calcium inserts with concentrations of 50/100/300 mg/ml were placed inside the solid HEwater phantom base.

Data was acquired on both PCCT and EID-CT single-energy mode with 120kVp, 50/100/200/400 mAs. For spectral image comparison, the same phantom was also scanned with the sequential dual-energy mode on EID-CT with radiation dose matched that of PCCT and EID-CT single-energy scans. Both PCCT and EID-CT images were reconstructed with 5 mm slice thickness, display FOV 240 mm, and 512x512 matrix size. The iodine and calcium concentration maps for PCCT and EID-CT dual-energy were both generated using the same in-house software tool, which utilized the 60/90 keV VMI pair. The root-mean-square-errors (RMSE) between measured concentrations and the ground-truth values across multiple concentration levels were calculated for each dose level.

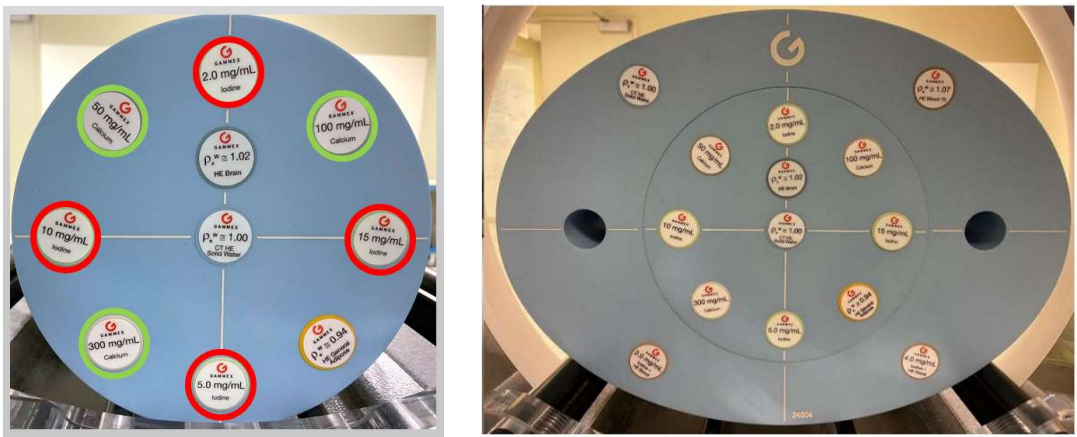

Figure 7: Pictures of the Multi-energy CT phantom (left: head size, right: body size) with different concentrations of iodine/calcium inserts used for quantitative spectral imaging evaluations.

**Noise impact of number of energy bins used in material decomposition**

To study the noise impact from number of energy bins used in material decomposition, a Multi-energy CT phantom (Sun Nuclear Corporation, WI, USA) of body size (40 cm x 30 cm) (Figure 7 right) was used. Data was acquired on PCCT with 120 kVp at 400 mAs. In this study, the default five energy bin data (30/45/55/65/80 keV) was recombined into two energy bins (30/65 keV) in NR mode, while the rest of the data processing and image reconstruction remain the same. Both 5-bin and 2-bin images were reconstructed with 5 mm slice thickness, display FOV 450 mm, and 512x512 matrix size. Two different ROIs were selected for noise comparison.

**Contrast-to-noise ratio**

To evaluate the image CNR of iodine and calcium in both counting and spectral modes, the aforementioned head-size Multi-energy CT phantom data was used. Image contrast levels were computed from ROIs at various iodine and calcium pins and ROI at HEwater background, where image noise was computed from the background ROI. For NR counting mode, we compared CNRs among EID-CT single-energy mode, original PCCT, and PCCT with matched MTF with EID-CT single-energy. For NR spectral mode, we compared CNRs between EID-CT dual-energy mode and PCCT with EDI-CT matched MTF at 70 keV VMI.

Furthermore, we estimated potential iodine contrast load reduction by PCCT when achieving the same CNR as EID-CT dual-energy mode at 50 keV. In particular, we first obtained CT number curves using linear regression of the four measured iodine CT numbers from 50 keV PCCT and EID-CT VMIs, since 50 keV VMI was typically used in clinical practice for iodine contrast study. Then we computed CNR as a function of iodine concentration, using the estimated iodine value and measured background value. At a given CNR, we found the corresponding iodine concentrations from EID-CT and PCCT separately, from which we computed how much less iodine PCCT needs as compared with EID-CT while achieving the same CNR in 50 keV VMI.

To evaluate the image CNR of low contrast objects, a Catphan CTP515 module (The Phantom Laboratory, NY, USA) (Figure 8) was scanned on both PCCT system and EID-CT system with single-energy mode (120kVp, 400mAs). Both PCCT and EID-CT images were reconstructed with 5 mm slice thickness, display FOV 200 mm, and 512x512 matrix size. To evaluate the image CNR, two small ROIs were selected at the 15 mm diameter target of 1.0% supra-slice region and the background to calculate the contrast, where the image noise was measured using the background ROI. The CNR values were compared among EID-CT single-energy, original PCCT NR counting, and PCCT NR counting with matched MTF with EID-CT.

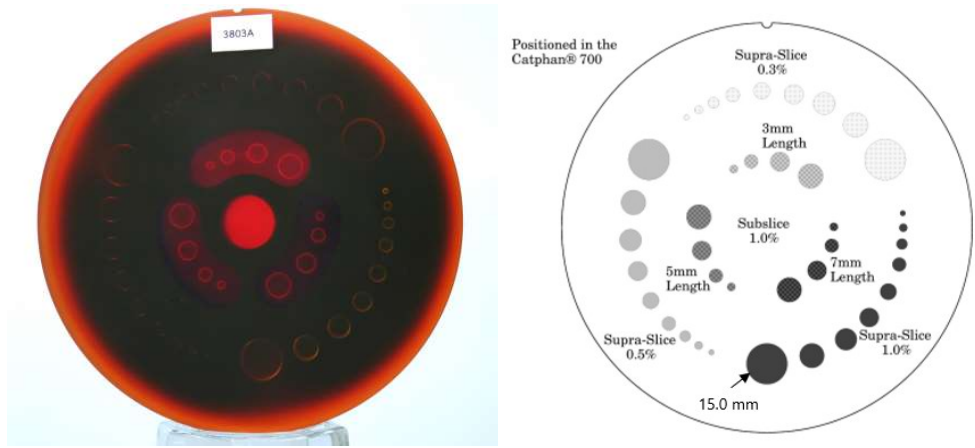

Figure 8: Catphan CTP515 (left) and rods configuration (right) (The Phantom Laboratory, NY, USA) used for image CNR evaluation.

## 3. RESULTS

The result section is organized as follows. We first present the evaluation results for PCCT system in the following order: NR counting mode, NR spectral mode, and SHR counting mode. Then we presented the comparison between PCCT system and conventional EID-CT system, where we first compared PCCT NR counting with EID-CT single-energy mode and then compared PCCT NR spectral with EID-CT dual-energy mode.

**3.1 PCCT Evaluation: NR Counting Mode**

First, a set of water images at different dose levels are displayed in Figure 9. All the images look uniform under a display window of WW/WL=300/0 HU and free of any noticeable artifacts. The mean CT number and noise as in the ROI standard deviation (SD) of the center FOV are displayed in Figure 10. For all the phantom sizes and dose levels, the CT number accuracy is mostly within ±3 HU. For different phantom sizes, one can observe that the measured noise-to-dose curves nicely follow the theoretical relationship: noise (SD) $\propto 1/\sqrt{dose}$ with Poisson distribution. This is a result of effectively removing the electronic noise in the PCD measurement and a good control of pulse pileup effect in the relevant flux range.

The image uniformity measurement results are displayed in Figure 11. The uniformities of the mean HU in the selected five ROIs are all within ±5 HU up to 32 cm water phantom, meeting the typical criteria for clinical diagnosis [31].

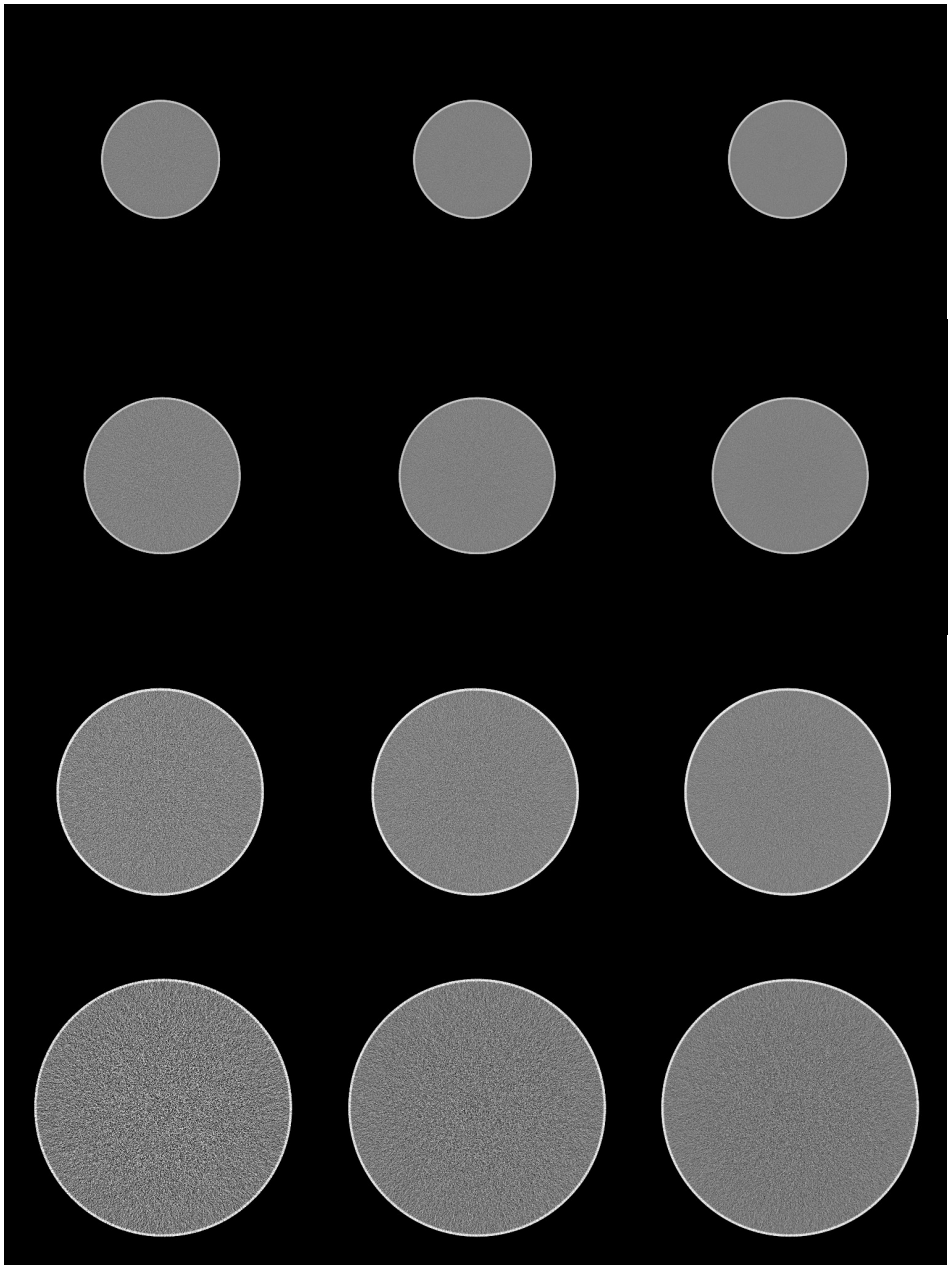

Figure 9: PCCT NR counting images of four different sizes of water phantoms. From top to bottom rows are: 18, 24, 32, and 40 cm diameter water phantoms respectively. From left to right columns are with 50, 100, and 200 mAs. Display window: WW/WL = 300/0 HU.

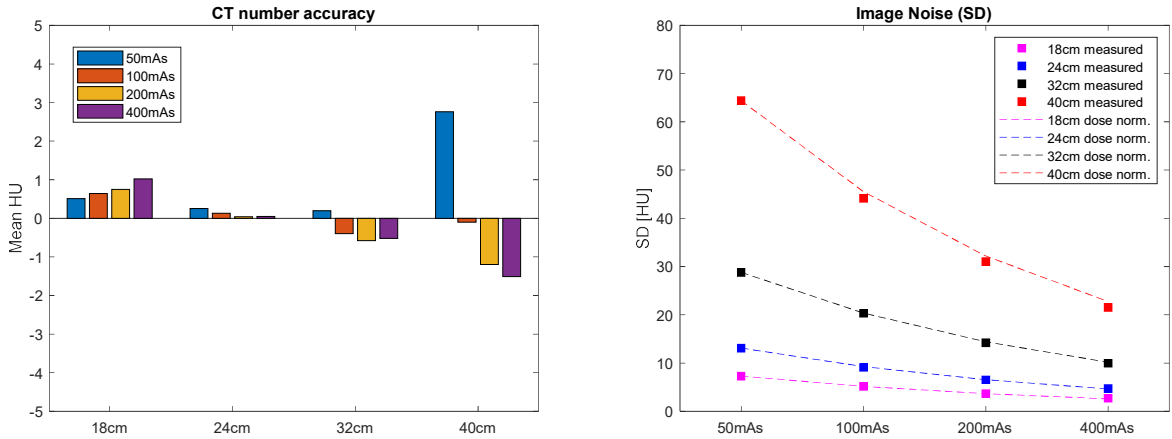

Figure 10: (Left) CT number (HU) and (right) noise standard deviation (SD) measurement from the center ROI of the PCCT NR counting water images. Most of the measurements are well within ±3HU across the phantom sizes and dose levels. (right) Image noise as measured in SD of the chosen ROI decreases linearly with $1/\sqrt{dose}$ (dashed lines) for all four phantom sizes.

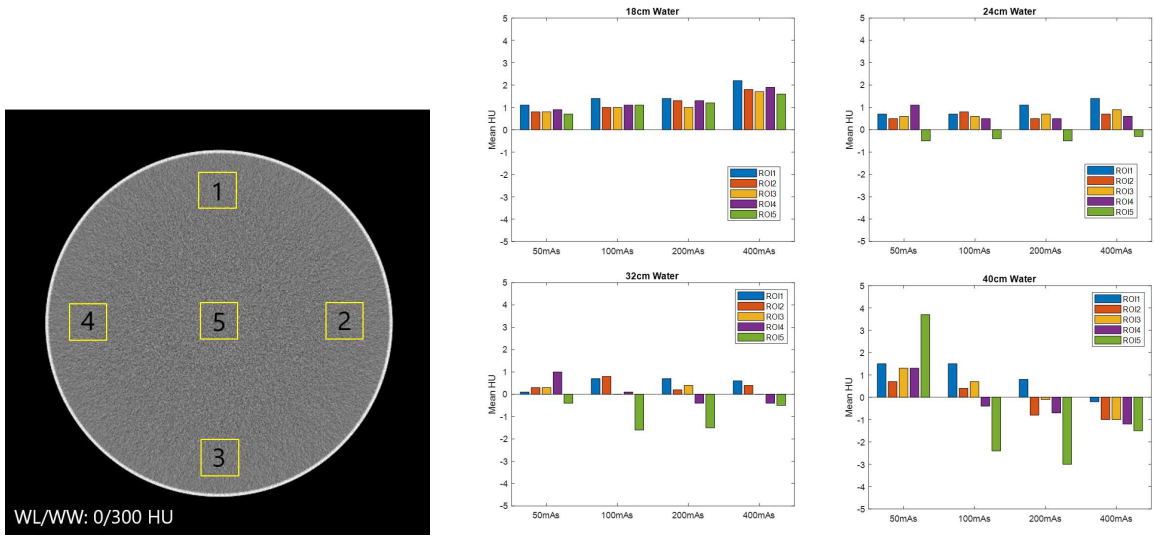

Figure 11: Five ROIs (left) were selected from the center and the periphery of water images (400 mAs, 40 cm water phantom, 5 mm slice thickness) for PCCT NR counting image CT number uniformity measurement. For all phantom sizes and dose levels, the uniformity is well within ±5 HU.

Evaluation results of CT number accuracy and consistency are shown in Figure 12 and Table 2. The mean CT numbers of the selected ROIs from PCCT images of the TOS phantom match well with typical EID-CT measurement range. PCCT results of all four dose levels also have good consistency with maximum deviations within ±1 HU.

Table 2: Canon TOS phantom mean CT number measurements from PCCT and EID-CT. The CT numbers of all five ROIs from PCCT images over four dose levels match well with EID-CT measurement.

| ROI# | EID Ref. CT number | PCCT CT number (mean) |
|---|---|---|
| (1) Water | 0±5 | -0.175 |
| (2) Delrin | 340±10 | 333.6 |
| (3) Acryl | 130±10 | 123.0 |
| (4) 66Nylon | 100±10 | 92.7 |
| (5) P.P. | -105±10 | -108.6 |

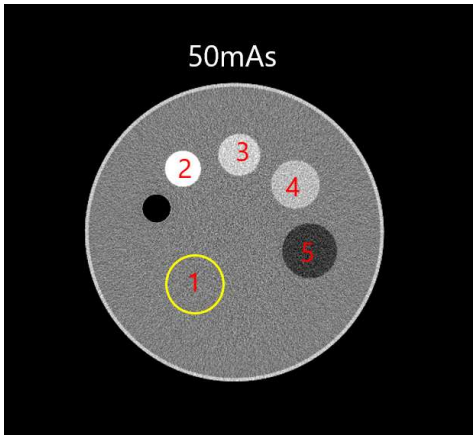
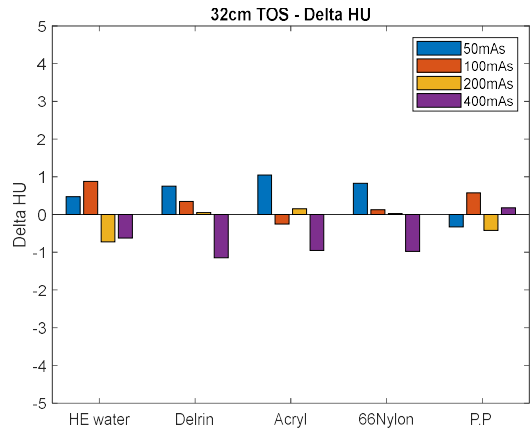

Figure 12: CT number measurements in a Canon TOS phantom PCCT image. All five ROIs selected at different materials have good CT number consistency across the dose levels. (Left): TOS phantom PCCT NR counting image. (Right): CT number deviation (HU) from the mean value of four measured dose levels.

The in-plane spatial resolution of the PCCT NR mode images was measured using the high contrast Teflon pin in Catphan CTP682 (Figure 13). With a large focal spot size and standard body kernel (FC13), the measured 50% modulation transfer function (MTF) is 0.33 lp/mm and the 10% MTF is 0.69 lp/mm (Figure 13).

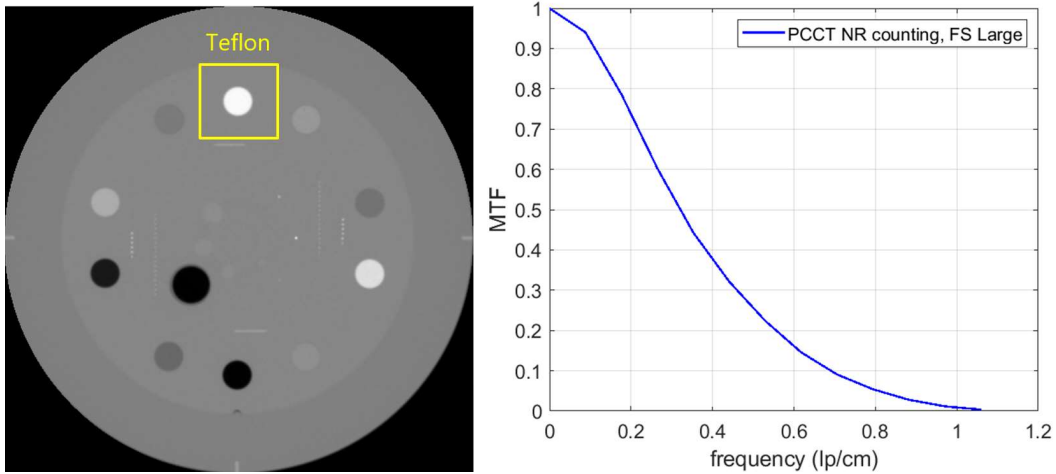

Figure 13: (Left) PCCT NR counting image of Catphan CTP682. The radial edge profile of the Teflon pin was used for the MTF measurement. (Right) For PCCT NR counting with FS Large, the 50% MTF is 0.33 lp/mm and the 10% MTF is 0.69 lp/mm.

**3.2 PCCT Evaluation: NR Spectral Mode**

The basic image quality was also evaluated for the PCCT NR VMIs generated from spectral mode with five energy bin output. Examples of a 24 cm water phantom VMIs are shown in Figure 14. The quantitative measurement in Table 3 shows that from low to high keV, VMIs have similar CT number uniformity as the counting image. In particular, the 70 keV VMI has very similar noise level as the counting image for this phantom size. The mean CT number of the center ROI and the CT number standard deviation were measured to assess the 70 keV VMI CT number accuracy for different sizes of water phantoms (Figure 15). For all the phantom sizes, the 70 keV VMIs have similar CT number accuracy and noise as the counting images (Table 4).

The in-plane spatial resolution of PCCT VMIs was also assessed and compared with PCCT NR counting (Figure 16). The results show that PCCT VMIs have consistent MTF as the PCCT NR counting image from low to high keV.

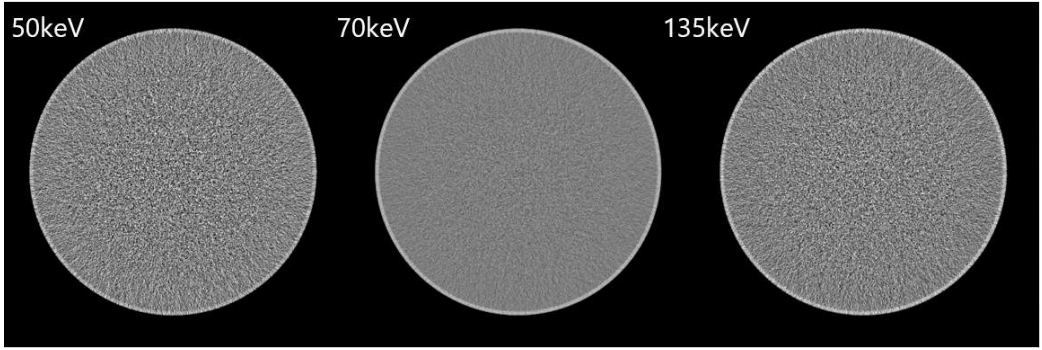

Figure 14: PCCT NR VMIs at 50/70/135 keV for a 24 cm water phantom (100 mAs, 0.62 mm slice thickness). Display window WW/WL: 400/0 HU.

Table 3: CT number uniformity measurement in 24 cm water phantom PCCT VMIs.

|  | ROI1 | ROI2 | ROI3 | ROI4 | ROI5 | Max ΔCT number |
|---|---|---|---|---|---|---|
| Counting | 1.1 | 1.2 | 0.8 | 0.8 | 0.0 | **1.2** |
| 50 keV | 2.9 | 3.8 | 1.9 | 4.2 | -1.5 | **5.7** |
| 70 keV | -2.6 | -1.2 | -2.2 | -2.0 | -2.4 | **1.4** |
| 135 keV | -0.8 | 0.3 | 0.3 | -0.6 | 2.4 | **3.2** |

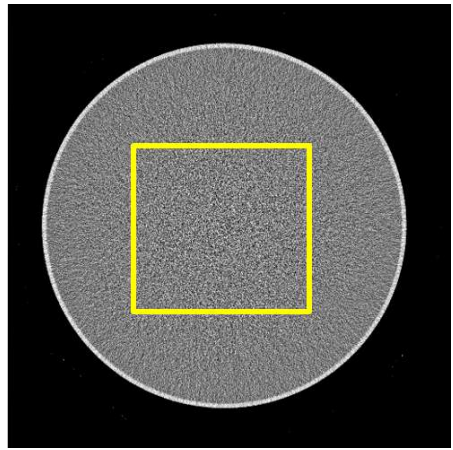

Figure 15: PCCT 70 keV VMI of a 32 cm water phantom (100 mAs, 0.62 mm slice thickness). Display window WW/WL: 400/0. The center ROI was selected for the CT number accuracy and noise comparison with NR counting images.

Table 4: PCCT NR counting and 70 keV VMI CT number accuracy and noise comparison for different sizes of water phantoms. The 70 keV VMI has similar noise as the counting image.

| Center ROI Mean/SD [HU] | Water 18 cm | Water 24 cm | Water 32 cm |
|---|---|---|---|
| Counting | -0.1/15.9 | 0.2/28.19 | 0.2/60.6 |
| 70 keV | -0.5/17.6 | -2.6/29.4 | 0.1/63.3 |

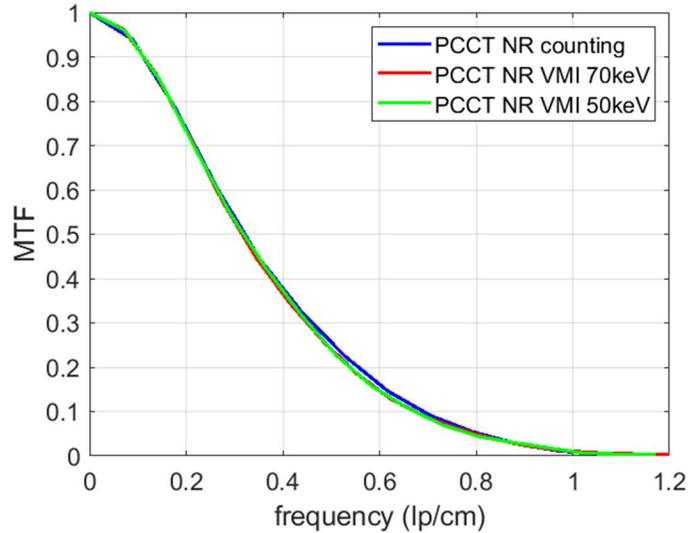

Figure 16: MTF measurement of PCCT NR counting and PCCT NR VMIs at 70 and 50 keV.

Compared to the counting image, one advantage of the VMIs is that the beam hardening artifact can be mostly removed given an accurate material decomposition. This is demonstrated in Figure 17**Error! Reference source not found.** that a 70 keV PCCT VMI shows a further reduced beam hardening artifact compared with the counting image of the Multi-energy CT head phantom. The minor shading between the high contrast rods in the counting image is further diminished in the 70 keV VMI. To quantify this improvement, the maximum CT number deviation of the neighboring ROIs was measured. The counting image has a 5.4 HU deviation while the 70 keV VMI has only 1.3 HU (Table 5).

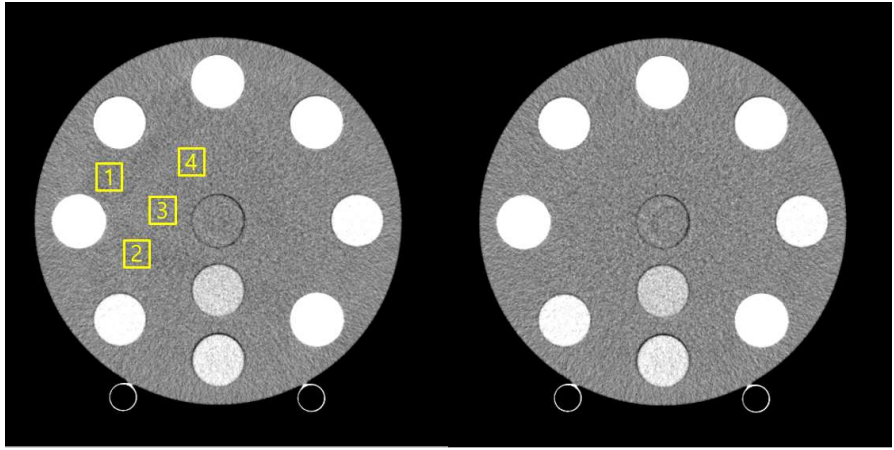

Figure 17: PCCT NR counting (left) and 70keV VMI (right) of the Multi-energy CT head phantom (100 mAs, 5mm slice thickness). Display window WW/WL:200/0 HU. 4 different small ROIs around the high contrast rods were selected to quantitatively assess the residual beam hardening artifact.

Table 5: Mean CT number within selected ROIs and maximum deviation for PCCT counting and 70 keV VMI of the Multi-energy CT head phantom. With highly attenuated rods, 70 keV VMI shows further improved CT number uniformity in the whole image than the counting mode due to reduced beam hardening artifact.

|  | Average CT number within ROI (HU) | | | | Max. ΔCT number |
|---|---|---|---|---|---|
|  | ROI1 | ROI2 | ROI3 | ROI4 |  |
| **Counting** | 0.2 | -0.1 | 5.3 | 2.4 | 5.4 |
| **70keV VMI** | 0.7 | 0.2 | -0.6 | -0.2 | 1.3 |

Figure 18 shows the VMIs of the Multi-energy CT head phantom along with the iodine and calcium maps. Table 6 and Table 7 present the measured iodine and calcium concentrations on the Multi-energy CT head phantom across multiple dose levels. Both iodine and calcium rods concentrations were estimated accurately for all concentration levels.

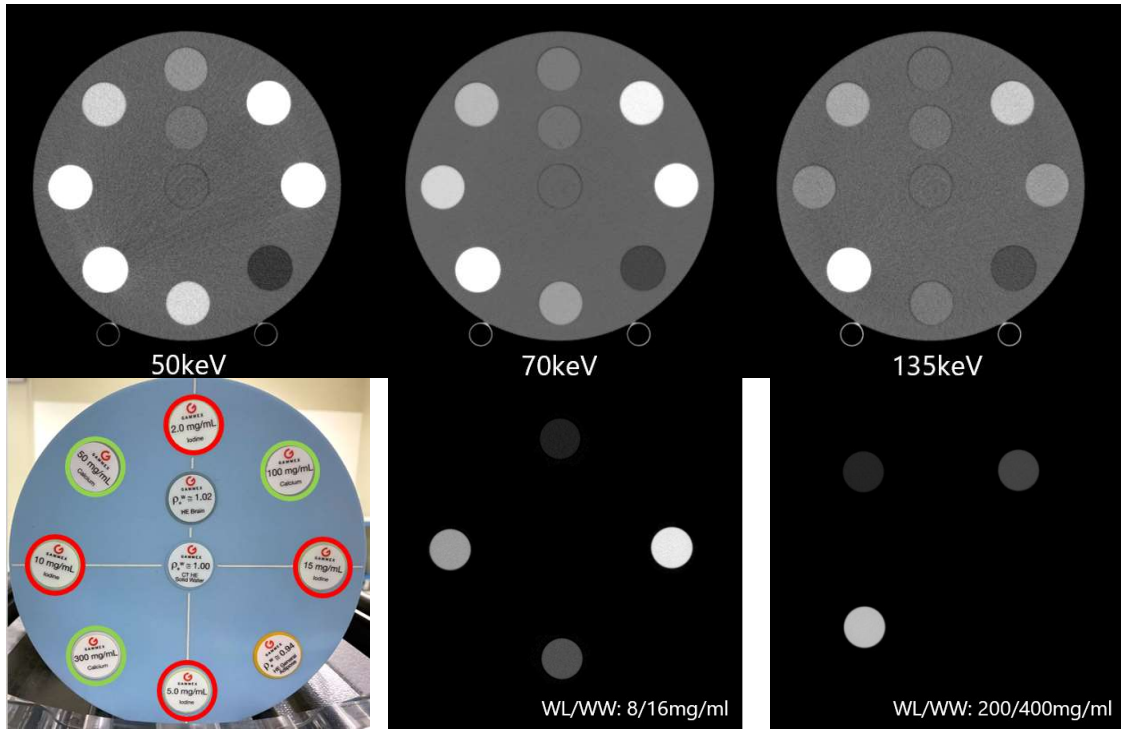

Figure 18: (Top) Multi-energy CT head phantom VMIs at 50/70/135 keV from PCCT scanned at 200 mAs. Display window WW/WL:600/60 HU. (Bottom) The picture of the phantom inserts arrangement (left). Iodine (middle) and calcium (right) maps generated from VMIs.

Table 6: Measured iodine concentrations from PCCT NR iodine map of the Multi-energy CT head phantom.

| Ground truth Iodine concentration (mg/ml) | PCCT NR spectral, measured mean and SD (mg/ml) | | | |
|---|---|---|---|---|
| | 50 mAs | 100 mAs | 200 mAs | 400 mAs |
| 2 | 2.02±0.29 | 2.01±0.21 | 2.03±0.14 | 2.09±0.11 |
| 5 | 5.07±0.32 | 5.04±0.21 | 5.04±0.15 | 5.11±0.12 |
| 10 | 9.91±0.30 | 9.85±0.23 | 9.80±0.16 | 9.90±0.13 |
| 15 | 14.85±0.34 | 14.76±0.23 | 14.61±0.19 | 14.76±0.14 |
| **RMSE** | **0.084** | **0.14** | **0.22** | **0.15** |

Table 7: Measured calcium concentrations from PCCT NR calcium map of the Multi-energy CT head phantom.

| Ground truth Calcium concentration (mg/ml) | PCCT measured mean and SD (mg/ml) | | | |
|---|---|---|---|---|
| | 50 mAs | 100 mAs | 200 mAs | 400 mAs |
| 50 | 57.69±2.87 | 57.57±2.08 | 57.63±1.46 | 58.00±1.15 |
| 100 | 106.88±3.12 | 106.57±2.20 | 106.43±1.63 | 107.01±1.23 |
| 300 | 302.58±3.95 | 301.14±3.00 | 297.50±1.95 | 298.80±1.46 |
| **RMSE** | **6.14** | **5.82** | **5.94** | **6.18** |

Figure 19 illustrates the comparison of 50 and 135 keV VMIs generated from a 5-energy bin decomposition and a 2-energy bin decomposition. From visual inspection, both sets of VMIs have good image quality in general, with the 2-energy bin decomposition image showing elevated noise and streak artifact between heavy inserts. We selected two ROIs in the background material for a quantitative comparison. The noise SD of the selected ROIs were measured at 50/70/135 keV VMIs for both cases, and the results are listed in Table 8. In this case, 50keV and 135keV VMIs with 2 energy-bin decomposition have ~6% to ~10% higher noise in the selected ROIs compared to the 5 energy-bin decomposition . For 70keV, the VMIs show similar noise.

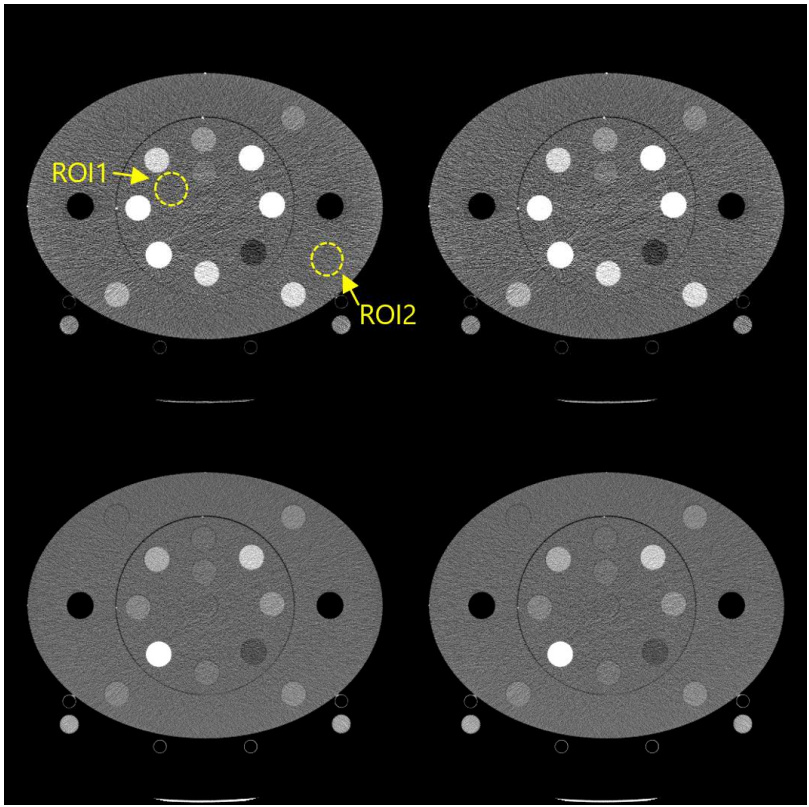

Figure 19: PCCT 50 keV (top row) and 135 keV (bottom) VMIs using 5-energy bin decomposition (left column) vs. 2-energy bin decomposition (right column) from the Multi-energy CT body phantom. Display window WW/WL: 600/60 HU. Two ROIs were selected for quantitative noise comparison.

Table 8: PCCT VMI noise with different energy bin data input for material decomposition from a 400 mAs Multi-energy CT head phantom scan. For all the reconstructed VMIs, 2-bin decomposition generates higher image noise in 50 keV and 135 keV than 5-bin decomposition.

| Material decomposition input | 50 keV SD ROI1/ROI2 | 70 keV SD ROI1/ROI2 | 135 keV SD ROI1/ROI2 |
|---|---|---|---|
| 5-bin (30/45/55/65/80 keV) | 70.9/59.7 | 16.1/13.9 | 37.9/32.2 |
| 2-bin (30/65 keV) | 77.2/65.7 | 16.2/14.7 | 41.2/34.2 |
| Noise increased in 2-bin decomposition | 8.9%/10.0% | 0.6%/5.7% | 8.7%/6.2% |

### 3.3 PCCT evaluation: SHR counting mode

The PCCT system allows for SHR imaging using the micro-pixel level readout for reconstruction. The NR counting and SHR counting images of the high-contrast line pair Catphan CTP714 are displayed in Figure 20**Error! Reference source not found.**. The 9 lp/cm bars can be visually identified in the NR image while at least 15 lp/cm pair can be identified in the SHR image. The MTF measurements of NR and SHR mode are displayed in Figure 21. At least a 91% improvement by SHR mode at 10% MTF is observed from the MTF curves. Although these results are still preliminary and do not represent the best SHR performance due to the relatively large focal size and suboptimal reconstruction kernel, the improvement of the spatial resolution is clearly demonstrated in the LUNGMAN phantom images in Figure 22. The SHR phantom image is much sharper and reveals more details of the structure.

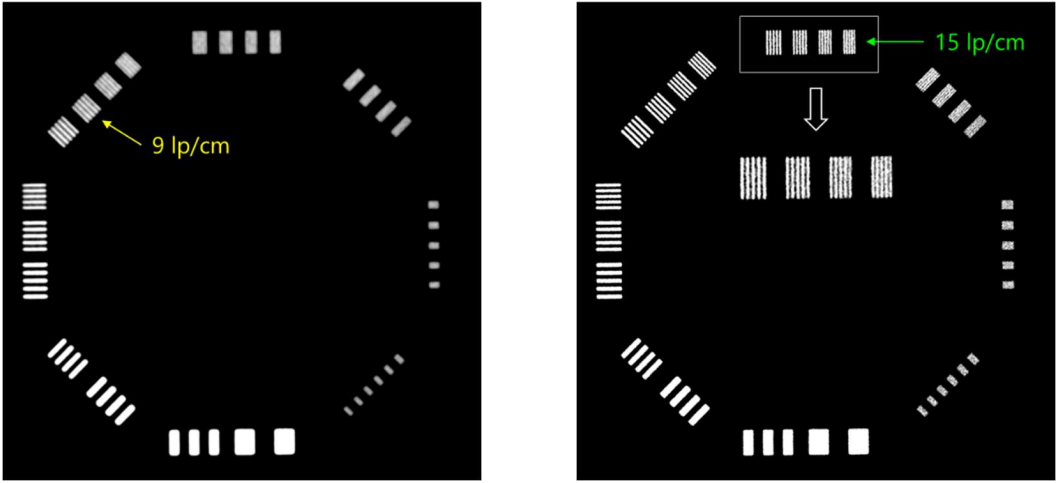

Figure 20: PCCT NR (left) and SHR (right) images of high contrast resolution bars in Catphan CTP714. A significant resolution improvement can be observed in SHR mode. Display window: [600 1400] HU.

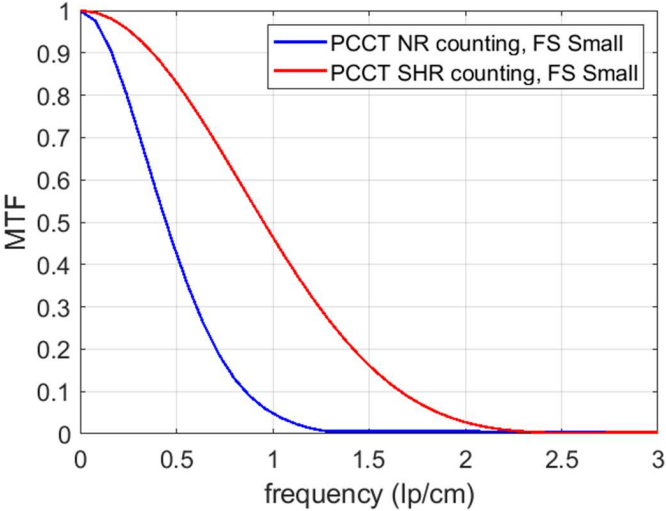

Figure 21: MTF measurement of PCCT NR and SHR modes. The MTF result shows that the PCCT SHR image has a 91% increase at 10% MTF over the NR image (1.65 lp/mm for SHR vs. 0.86 lp/mm for NR).

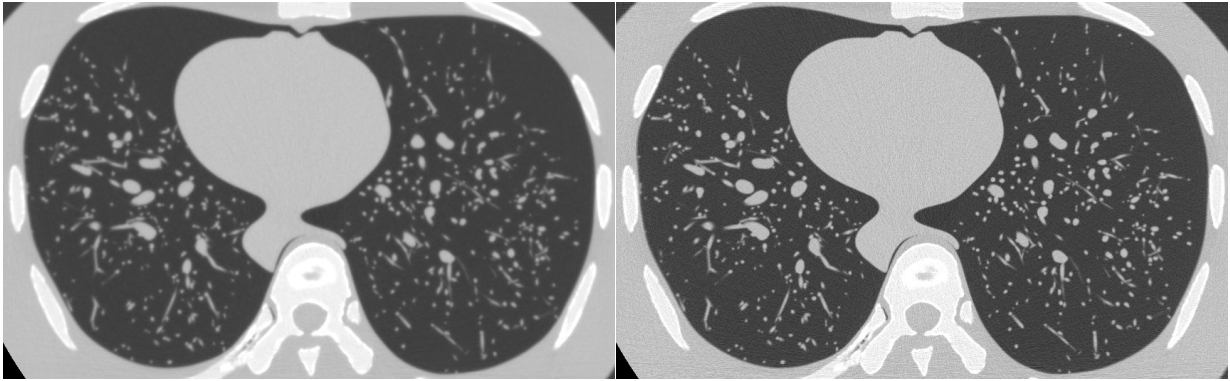

Figure 22: LUNGMAN phantom images of PCCT NR (left) and SHR (right) counting modes. With much smaller detector pitch, SHR mode produces much sharper image features than NR mode. Display window: [-1500 500] HU.

**3.4 Comparison of PCCT NR Counting Mode and EID-CT Single-Energy Mode**

Figure 23 illustrates the comparison of MTFs between EID-CT single-energy and PCCT NR counting, with and without matched MTF. For PCCT NR counting, the 50% MTF is 0.33 lp/mm and the 10% MTF is 0.69 lp/mm, which is higher than EID-CT single-energy (0.30 lp/mm at 50% MTF and 0.60 lp/mm at 10% MTF.) Compared with EID-CT single-energy, PCCT NR counting produces 15% improvement of the MTF.

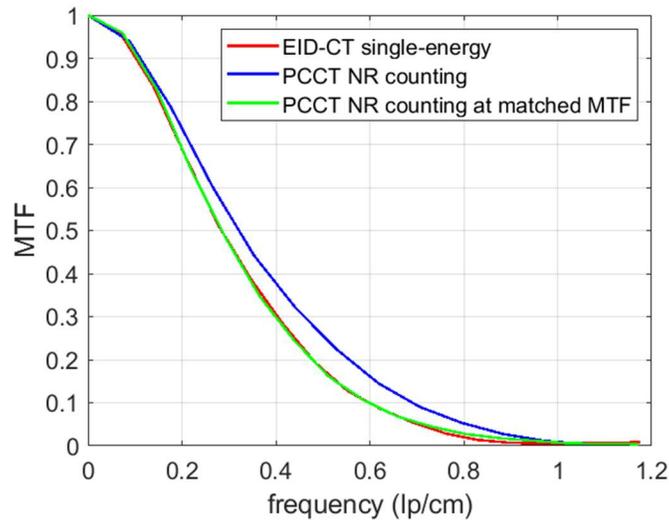

Figure 23: MTF measurements of EID-CT single-energy, PCCT NR counting with and without matched MTF. With additional smoothing on the PCCT image, the MTF can be matched with the EID-CT for noise comparison.

To further demonstrate the potential benefit of PCCT, we compared the image noise by matching the MTF of PCCT NR counting image with that of EID-CT single-energy image. The matched MTFs are shown in Figure 23. Figure 24 illustrates the 40cm water phantom images at 50 mAs from EID-CT single-energy and PCCT NR counting. PCCT NR counting images show improved uniformity over the entire phantom while the EID-CT has noticeable CT number bias towards the center (~14 HU). The noise grain of the PCCT image is also finer, indicating that the noise power spectrum shifts toward higher frequency. Figure 25 illustrates the image noise measured on the 40cm water phantom across multiple dose levels. Results show that original PCCT NR counting images have lower noise than EID-CT single-energy images at all dose levels, where the noises reduction is more significant as dose decreases. After matching MTF, the PCCT images show significant noise reduction for all dose levels, up to 53% at 50 mAs, which is equivalent to 78% dose reduction when matching the image noise as EID-CT.

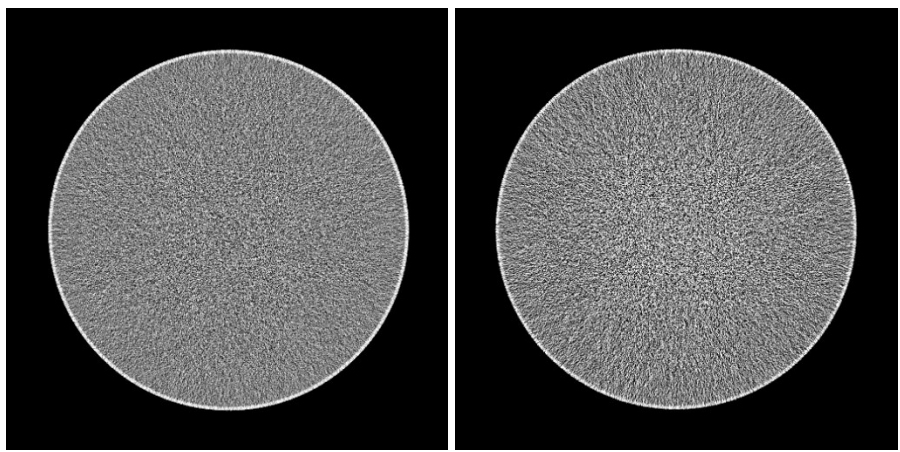

Figure 24: Images of a 40 cm water phantom acquired at 50 mAs from EID-CT single-energy (left) and PCCT NR counting mode (right). Display window WW/WL: 200/0 HU. PCCT NR counting image show improved uniformity over the entire phantom as compared to EID-CT.

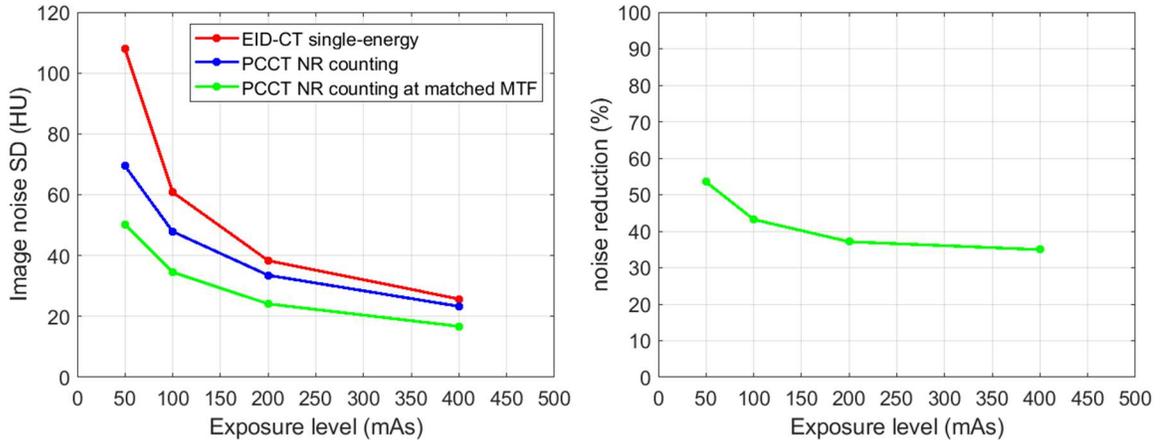

Figure 25: (Left) Noise SD on a 40 cm water phantom from 50 mAs to 400 mAs. (Right) Noise reduction of PCCT NR counting compared with EID-CT single-energy at matched MTF. After MTF matching with EID-CT, PCCT images have much greater noise reduction as compared to EID-CT single-energy across all dose levels.

Figure 26 shows the comparison of image CNR for iodine and calcium objects between EID-CT single-energy and PCCT NR counting before and after matching MTF with EID-CT. For the PCCT NR counting mode, the original images show higher CNR than EID-CT for all the evaluated iodine and calcium concentrations, thanks to more optimal photon weighting and reduced noise in PCCT. The MTF-matched images show additional improvement due to further reduced image noise. Both the original PCCT counting and MTF-matched PCCT counting present higher CNR than EID single-energy CT in iodine and calcium objects across multiple concentrations. The average CNR improvement over EID-CT single-energy is 7% for original PCCT NR counting and 31% for MTF-matched PCCT NR counting.

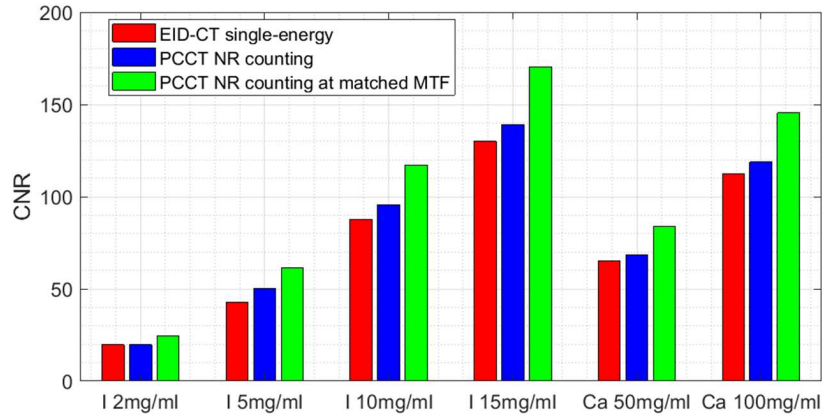

Figure 26: CNR comparison among EID-CT single-energy, PCCT NR counting before and after matching MTF with EID-CT. Both the original PCCT counting and MTF-matched PCCT counting present higher CNR than EID single-energy CT in iodine and calcium objects across multiple concentrations.

Figure 27 shows the Catphan CTP515 images for EID-CT single-energy and PCCT NR counting image before and after matching MTF with EID-CT. The original PCCT NR counting image has better edge definition for all the low contrast objects and visually allows one to better identify smaller objects with lower contrast from the background. The CNR measurements are summarized in Table 9. With matched MTF, the PCCT image increases the low contrast CNR from 2.5 to 4.6, a ~84% improvement over EID-CT. The improved spatial resolution in the PCCT image provides better edge definition for those low contrast inserts, hence could increase the detectability of such low contrast objects.

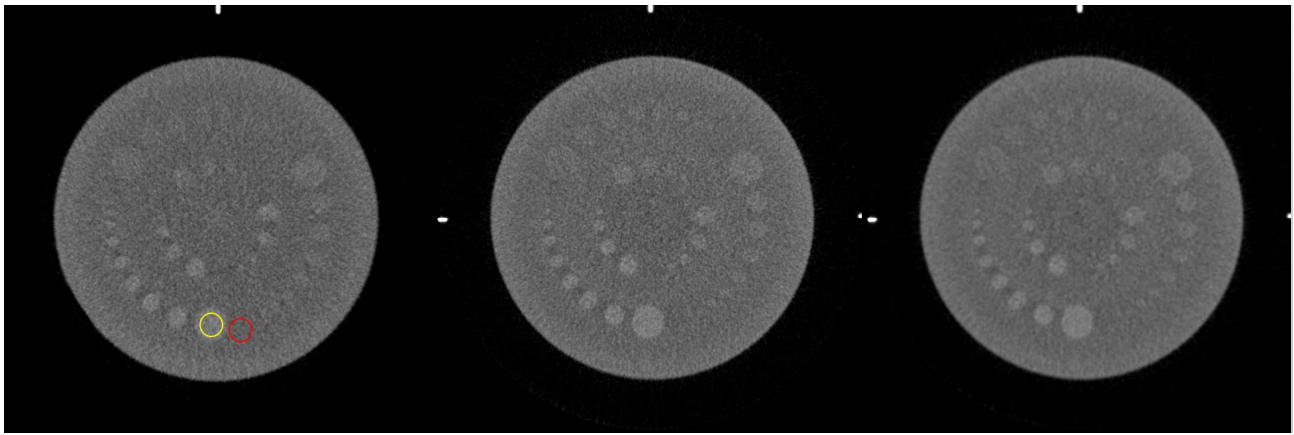

Figure 27: Images of Catphan CTP515 from EID-CT (left), original PCCT NR counting (middle), and PCCT NR counting image with MTF matched with EID-CT (right). Display window: WW/WL=100/60 HU. ROIs used to compute CNR are illustrated in the EID-CT image (left). PCCT NR counting images show better edge definition for low contrast inserts than EID-CT single-energy image.

Table 9: Low contrast object CNR comparison between PCCT NR counting and EID-CT single-energy using Catphan CTP515. CNR of the 15 mm diameter target at 1.0% contrast level increases from 2.5 in EID-CT to 4.6 in PCCT with matched-image MTF.

|  | EID-CT single-energy | PCCT NR counting | PCCT NR counting w/ matched MTF |
|---|---|---|---|
| Object | 60.2±3.6 HU | 58.5±3.2 HU | 58.5±1.8 HU |
| Background | 50.6±3.9 HU | 48.3±3.9 HU | 48.3±2.2 HU |
| Contrast | 9.6 | 10.2 | 10.2 |
| CNR | 2.5 | 2.6 | 4.6 |

**3.5 Comparison of PCCT NR Spectral Mode and EID-CT Dual-Energy Mode**

Figure 28 illustrates the comparison of in-plane spatial resolution between PCCT NR VMIs and EID-CT dual-energy VMIs. Results show that PCCT VMIs have MTFs superior to those of EID-CT dual-energy VMIs, especially at 50 keV with ~20% improvement at 10% MTF. Figure 29 (left) illustrates the comparison of VMI noise between EID-CT dual-energy and PCCT NR spectral before and after MTF matching, measured using the head-size Multi-energy CT phantom. Before MTF matching, the original PCCT VMIs have lower noise across the whole energy range. By trading off that additional spatial resolution, the noise is further reduced in MTF-matched PCCT NR VMIs. Another observation is that the lowest noise of PCCT VMIs is in the vicinity of 65 keV for this phantom and is consistent with the counting image noise. This can be expected since the same photon statistics are used for both counting and spectral mode processing, and could indicate that little additional noise is introduced during the PCCT material decomposition step.

With the improved noise and MTF in VMIs, we compare the material quantification results between PCCT NR spectral and EID-CT dual-energy at matched MTF at 70 keV with matched radiation dose (Table 10). The overall quantification accuracy of PCCT NR spectral is improved for all concentrations with significantly reduced noise as compared with EID-CT dual-energy.

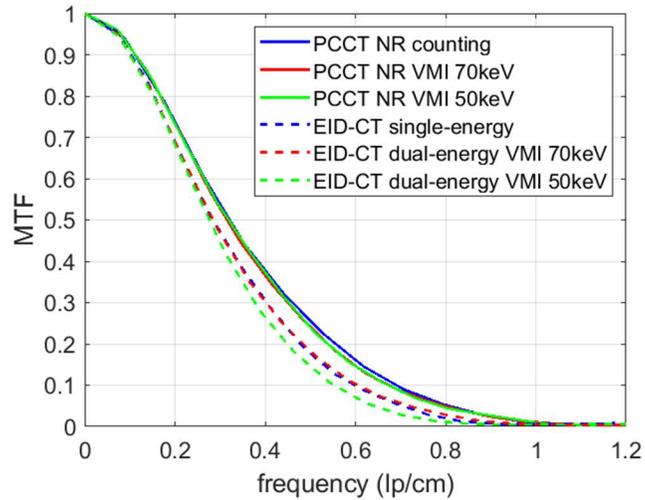

Figure 28: MTF measurement of PCCT NR counting and PCCT NR VMIs at 70 and 50 keV, compared with EID-CT single-energy and EID-CT dual-energy VMIs at 70 and 50 keV. PCCT VMIs have MTFs superior to those of EID-CT dual-energy VMIs.

Figure 29 (right) illustrates the ratio of PCCT VMI CNR to EID-CT VMI CNR at matched MTF from 40 to 70 keV. Result shows that PCCT spectral produces consistently higher CNR than EID-CT dual-energy, with on average a 56% improvement and a minimum of 26% improvement (2 mg/ml at 55 keV). One thing to note is that, even with matched 70 keV VMI resolution, the PCCT 50 keV VMI still has superior resolution compared to EID-CT (Figure 28), providing room for further noise-resolution trade-off to optimize the low keV VMI diagnosis capability.

At the same CNR for both PCCT NR spectral and EID-CT dual-energy at 50 keV, we estimated how much iodine load reduction can be achieved in PCCT NR spectral. Figure 30 (left) shows the estimated CNR of iodinated object as a function of iodine concentration for both PCCT NR spectral and EID-CT dual-energy. Figure 30 (right) shows that, from 2 mg/ml to 15 mg/ml, an average of ~32% less iodine contrast load can be used in PCCT to achieve same CNR in EID-CT dual-energy at 50keV VMI.

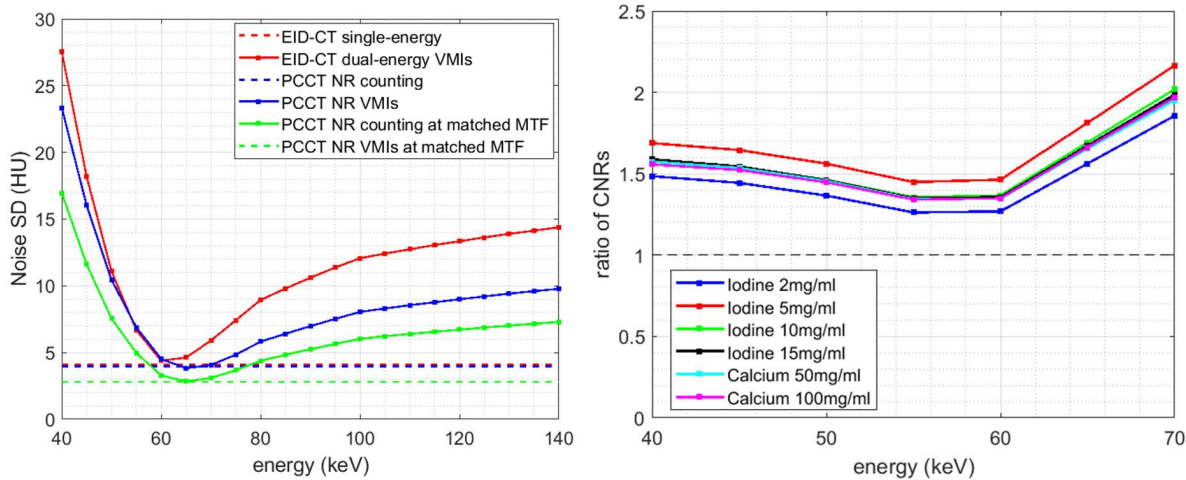

Figure 29: (Left) Comparison of image noise measured on multiple imaging modes. (Right) Ratio of MTF-matched PCCT VMI CNRs to EID-CT VMI CNRs for different materials and concentrations.

Table 10: Iodine concentration estimation comparison between PCCT NR spectral and EID-CT dual-energy with matched MTF at 70keV, using 100 mAs scans of the Multi-energy CT head phantom.

| Ground Truth iodine concentration (mg/ml) | PCCT NR spectral at matched MTF | EID-CT dual-energy |
|---|---|---|
| 2 | 2.01±0.15 | 1.83±0.27 |
| 5 | 5.05±0.15 | 4.33±0.25 |
| 10 | 9.85±0.17 | 9.27±0.23 |
| 15 | 14.77±0.17 | 14.22±0.25 |
| **RMSE** | **0.14** | **0.64** |

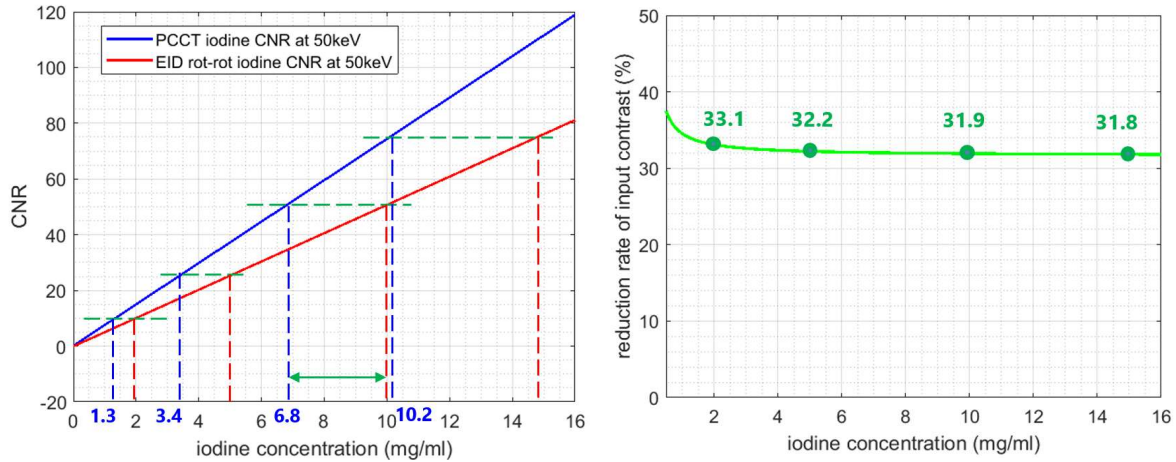

Figure 30: (Left) Estimated iodine CNR as a function of iodine concentration for both PCCT NR spectral and EID-CT dual-energy. For all the concentration levels, PCCT needs significantly less iodine contrast concentration to achieve the same CNR at 50 keV VMI as measured in EID-CT. (Right) Reduction rate of iodine contrast load by PCCT as compared with EID-CT when achieving same CNR at 50 keV.

## 4. DISCUSSION AND SUMMARY

In this work, we introduced our first prototype full field-of-view photon counting CT system and evaluated its imaging performance through phantom studies. By comparing with EID-CT performance in the same tasks, we demonstrated that the prototype system is capable of producing diagnostic quality images at all assessed clinical dose levels with superior performance in the following aspects:

1. PCCT significantly reduces image noise, particularly in low dose cases (Figure 25). This is mainly due to the removal of electronic noise in the counting mode with an appropriate energy threshold [5][9]. We observed up to 37% noise reduced in a 40 cm water phantom counting image with the same scan settings as EID-CT at 50 mAs, and at different dose levels, the noise nicely follows Poisson statistics (Figure 10). Another associated benefit is fewer image artifacts from photon starvation, which potentially allows for lower dose to achieve the same image quality as EID-CT in the clinical setting (Figure 24).

2. PCCT improves image spatial resolution. With the same in-plane detection pitch as the EID-CT, PCCT NR mode still demonstrate better MTF. We observed a ~15% increase in 10% MTF (Figure 23) compared to EID-CT which is mainly due to the reduced crosstalk between neighboring pixels. In PCCT, the crosstalk effect is mainly from the charge-sharing effect, and is mostly confined between neighboring micro-pixels [34]. While for EID-CT, due to the two-stage conversion process, the crosstalk kernel is at least at the width of a macro-pixel with larger magnitude [13]. PCCT can also have more flexible readout modes, from the current '3×3' grid to other types of summing grids to further optimize the NR mode resolution. With improved spatial resolution, it provides more room for trade-off when applying denoising to optimize the image quality depending on the tasks. As an example, our results demonstrated that up to 53% noise reduction can be achieved by simply matching the MTF with EID-CT for a 40 cm water phantom at 50 mAs, which is equivalent to a 78% dose reduction (Figure 25).

3. PCCT can produce SHR images from the micro-pixel level readout. With roughly 1/9 of the conventional detector pixel size, the limiting resolution is largely increased and images are much sharper and reveal many more details. We observed a 91% increase in 10% MTF (Figure 21) from the NR to SHR with a standard body recon kernel (FC13), and it is yet the full potential of the current PCCT due to the relatively large focal spot size and suboptimal reconstruction kernel.

4. PCCT can enable spectral imaging without introducing additional complications from workflow or data temporal inconsistency like conventional DECT using dual source or fast kVp switching [13][14]. Using a projection domain decomposition approach, it can flexibly utilize multiple (≥2) energy bin input with more spectral information and further reduce spectral image noise. It is demonstrated that PCCT VMIs using 5-energy bins have better quantitative accuracy and much lower noise than EID rotate-rotate DECT (Table 10). The resulting material quantification has excellent accuracy with a mean RSME of 0.15 mg/ml and 6.02 mg/ml for iodine and calcium (Table 6 and Table 7), respectively. It is also demonstrated that PCCT NR VMIs further reduce beam hardening artifacts and well preserve the same spatial resolution as the NR counting mode (Figure 28). We have demonstrated that the VMIs generated with 5-energy bin data have lower image noise at both low and high keV range than the 2-energy bin data (Table 8), mostly due to the additional spectral information [32][33]. Comparing to the image-based material decomposition approach, which requires reconstruction of individual energy bin images in the first place (more prone to the beam hardening artifacts and detector response modeling errors when energy bins get finer) [9][32] , one advantage of using projection domain decomposition is to handle multiple energy bin data more coherently, hence the results could be more resilient to the potential PCD forward modeling errors and the statistical noise in the individual bin measurement. These findings may open new opportunities in spectral imaging applications and provide new insights for vendors to optimize the data usage.

5. PCCT registers each photon equally, and together with the reduced noise, PCCT can generate better CNR in different tasks. As we evaluated through this work, in both high contrast and low contrast cases, PCCT demonstrated superior performance in CNR for both counting and spectral modes and can potentially reduce the iodine contrast by 32% (Figure 26) while maintaining the same CNR as EID-CT for diagnosis.

There are some limitations in our study. The current prototype system covers up to about 1 cm in Z at isocenter and only supports circular scans at 1 second per rotation. It does not support real time data processing and reconstruction, and cannot conduct more complex scan series to fully mimic clinical workflows. It is also not equipped with high precision tube with high precision focal spot to fully demonstrate the SHR mode spatial resolution capability, or more advanced features such as tube current modulation (AEC) and AI-based denoising for additional dose saving and image quality enhancement. These limitations are mostly engineering related and will be gradually resolved in future development. Therefore, the results we report here do not represent the full performance of the future product.

In conclusion, we performed comprehensive phantom imaging evaluations on our first CdZnTe-based prototype PCCT system. Through rigorously designed experiments and analysis, the initial results demonstrate multiple advantages over conventional EID-CT and provide us in-depth understanding of the current PCD performance under clinical scan conditions and invaluable insights on design trade-offs. Currently, a new generation of clinical prototype PCCT system is under development. With a wider Z coverage, a high precision X-ray tube, faster rotation speeds and improved scan workflows, it will enable patient studies to further demonstrate the clinical values with photon counting detection technology.

## REFERENCES

[1] Taguchi, K., Iwanczyk, JS. Vision 20/20: single photon counting x-ray detectors in medical imaging. Med Phys 2013;40(10):100901.
[2] Kappler, S., Henning, A., Kreisler, B., Schoeck, F., Stierstorfer, K., Flohr, T. Photon counting CT at elevated x-ray tube currents: contrast stability, image noise and multi-energy performance. Proc. of SPIE: medical imaging 2014, physics of medical imaging. Vol 9033.
[3] Persson, M., Huber, B., Karlsson, S., et al. Energy-resolved CT imaging with a photon-counting silicon-strip detector. Phys Med Biol 2014;59(22):6709–6727.
[4] Gutjahr, R., Halaweish, AF., Yu, Z., et al. Human imaging with photon counting based computed tomography at clinical dose levels: contrast-to-noise ratio and cadaver studies. Invest Radiol 2016;51(7):421–429.

[5] Danielsson et al. Photon-counting x-ray detectors for CT. Phys Med Biol 2021 Jan 29;66(3):03TR01.

[6] FDA news release: FDA clears first major imaging device advancement for Computed Tomography in nearly a decade. 2021 Sep 30.

[7] Pourmorteza, A., Symons, R., et al. Photon-counting CT of the bain: in vivo human results and image-quality assessment, AJNR Am J Neuroradiol. 2017 Dec; 38(12):2257-2263.

[8] Rajendran, K., Petersilka, M., Henning, A., et al. First clinical photon-counting detector CT System: technical evaluation. Radiology 2022; 000:1-9.

[9] Willemink, M., Persson, M., Pourmorteza, A., Pelc, N., Photon-counting CT: technical principles and clinical prospects. Radiology 2018; 00:1-20.

[10] Leng, S., et al. Photon-counting detector CT: system design and clinical applications of an emerging technology. RadioGraphics 2019; Vol.39, No. 3.

[11] Rajendran, K., Petersilka, M., et al. Full field-of-view, high-resolution, photon counting detector CT: technical assessment and initial patient experience. Phys. Med. Biol. 2021 Oct 27;66(2):10.

[12] Si-Mohamed, S., Boccalini, S., et al. Coronary CT angiography with photon-counting CT: first-in-human results Radiology 2022 May;303(2):303-313.

[13] Shefer, E., Altman, A., Behling, R., et al. State of the art of CT detectors and sources: a literature review. Curr Radiol Rep 2013; 1(1):76-91.

[14] Johnson, T., Fink, C., et al. Dual energy CT in clinical practice. Heidelberg, Germany: Springer, 2011.

[15] McCollough, CH., Leng, S., Yu, L., Fletcher, JG. Dual- and multi-energy CT: principles, technical approaches, and clinical applications. Radiology 2015;276(3):637-653.

[16] Roessl, E. and Proksa, R., "K-edge imaging in x-ray computed tomography using multi-bin photon counting detectors," Physics in Medicine and Biology 52(15), 4679–4696 (2007).

[17] Si-Mohamed, S., Cormode, D. P., et al. Evaluation of spectral photon counting computed tomography K-edge imaging for determination of gold nanoparticle biodistribution in vivo. Nanoscale, 2017, 9, 18246-18257.

[18] Ostadhossein, F., et al. Multi-"color" delineation of bone micro damages using ligand-directed sub-5 nm hafnia nanodots and photon counting CT imaging. Advanced Functional Materials, 2020, 30, 1904936.

[19] Eisen, Y., Shor, A., Mardor, I. CdTe and CdZnTe gamma ray detectors for medical and industrial imaging systems, Nucl. Instr. And Meth. A, Vol 428, Issue 1, June 1999, 158-170.

[20] Muenzel, D., et al. Spectral photon-counting CT: initial experience with dual-contrast agent K-edge colonography. Radiology 2017, 283, 723-728.

[21] Blevis, I. X-ray detector for spectral photon counting CT. Spectral, photon counting computed tomography technology and applications, CRC Press

[22] Symon, R., Morris, J., et al. Coronary CT Angiography: variability of CT scanners and readers in measurement of plague volume. Radiology, Vol. 281, No. 3, 2016.

[23] Taguchi, K., Zhang, M., Frey, E., Wang, X., Modeling the performance of a photon counting x-ray detector for CT: energy response and pulse pileup effects. Med. Phys. 38 (2), Feb. 2011.

[24] Alvarez, R. E., Macovski, A., Energy-selective reconstructions in x-ray computerized tomography. Phys. Med. Biol. 1976 21 733-44.

[25] Roessl, E., Proksa, R., K-edge imaging in x-ray computed tomography using multi-bin photon counting detectors. Phys. Med. Biol. 52 (2007) 4679-4696.

[26] McCollough, C. et al, Principles and applications of multienergy CT: Report of AAPM Task Group 291, Med. Phys. Vol 47-7, 2020.

[27] Leng, S., Zhou, W., Yu, Z., et al. Spectral performance of a whole-body research photon counting detector CT: quantitative accuracy in derived image sets. Phys. Med. Biol. 2017;62(17):7216–7232.

[28] Liu, X., Persson, M., et al. Spectral response model for a multibin photon-counting spectral computed tomography detector and its applications, Journal of Medical Imaging 2(3), 033502 (2015).

[29] Schmidt, T. G., Barber, R. F., and Sidky, E. Y., A spectral CT method to directly estimate basis material maps from experimental photon-counting data, IEEE Transactions on Medical Imaging 36, 1808-1819 (2017).

[30] Dickmann, J., Maier, J., et al. A count rate-dependent method for spectral distortion correction in photon counting CT. Proc. Of SPIE Vol. 10573 1057311-1 (2018).

[31] IEC 61223-3-5: 2019

[32] Faby, S., Kuchenbecker, S., et al. Performance of today's dual energy CT and future multi energy CT in virtual non-contrast imaging and in iodine quantification: A simulation study. Med. Phys. 2015 Jul;42(7):4349-66.

[33] Alvarez, R., Estimator for photon counting energy selective X-ray imaging with multibin pulse height analysis, Med. Phys. 38(5), May 2011.

[34] Zhan, X., Wu, S., Hein, I., Markov, N., et al. A study of cross-talk effect in pixelated photon counting detectors and impact to system imaging performance. To be published in Proc. Of SPIE 2023.